\documentclass[preprint,prl,showkeys,showpacs,longbibliography]{revtex4-1}
\usepackage[utf8]{inputenc}
\usepackage{stmaryrd}
\usepackage{amsmath}
\usepackage{amssymb}
\usepackage{graphicx}
\usepackage{dcolumn}
\usepackage{bm}
\usepackage{multirow}
\usepackage[colorlinks,linkcolor=blue,citecolor=blue,urlcolor=blue]{hyperref}
\usepackage{color}
\usepackage{ulem}
\usepackage{comment}
\usepackage{gensymb}

\begin{document}
\title{Facet dependent surface energy gap on magnetic topological insulators}
\author{Hengxin Tan}
\author{Binghai Yan}
\email{binghai.yan@weizmann.ac.il}
\affiliation{Department of Condensed Matter Physics, Weizmann Institute of Science, Rehovot 7610001, Israel}

\date{\today}

\begin{abstract}

Magnetic topological insulators (MnBi$_2$Te$_4$)(Bi$_2$Te$_3$)$_n$ ($n=0,1,2,3$) are promising to realize exotic topological states such as the quantum anomalous Hall effect (QAHE) and axion insulator (AI), where the Bi$_2$Te$_3$ layer introduces versatility to engineer electronic and magnetic properties. However, whether surface states on the Bi$_2$Te$_3$ terminated facet are gapless or gapped is debated, and its consequences in thin-film properties are rarely discussed. In this work, we find that the Bi$_2$Te$_3$ terminated facets are gapless for $n \ge 1$ compounds by calculations. Despite that the surface Bi$_2$Te$_3$ (one layer or more) and underlying MnBi$_2$Te$_4$ layers hybridize and give rise to a gap, such a hybridization gap overlaps with bulk valence bands, leading to a gapless surface after all. Such a metallic surface poses a fundamental challenge to realize QAHE or AI, which requires an insulating gap in thin films with at least one Bi$_2$Te$_3$ surface. In theory, the insulating phase can still be realized in a film if both surfaces are MnBi$_2$Te$_4$ layers. Otherwise, it requires that the film thickness is less than 10$\sim$20 nm to push down bulk valence bands via the size effect. Our work paves the way to understand surface states and design bulk-insulating quantum devices in magnetic topological materials.

\end{abstract}
\maketitle

% \section{Introduction}
$Introduction.$
The magnetic topological quantum materials \cite{Smejkal2018,tokura2019} which combine the non-trivial band topology \cite{RevModPhys.82.3045,RevModPhys.83.1057} with magnetism display intriguing quantum phenomena, such as
quantum anomalous Hall effect (QAHE) \cite{Liu2008,science.1187485,chang2013}, axion insulator (AI) \cite{PhysRevB.78.195424,mogi2017magnetic,Xiao2018},
and Weyl semimetals \cite{Wan2011,yan2017,RevModPhys.90.015001}.
Recently, an intrinsic magnetic topological insulator
MnBi$_2$Te$_4$ (MBT) was discovered \cite{gong2019experimental,li2019intrinsic,PhysRevLett.122.206401,otrokov2019prediction} and attracted extensive interest \cite{wu2019natural,PhysRevLett.125.037201,PhysRevLett.125.117205,PhysRevX.11.031032,PhysRevLett.122.107202,lei2020magnetized,PhysRevLett.127.236402,WANG2021100098,Zhao2021}.

The MnBi$_2$Te$_4$ consists of seven atomic layers (called septuple layers, SL) in each van der Waals layer.
The SL exhibits ferromagnetic (FM) coupling between Mn atoms inside the layer and antiferromagnetic (AFM) coupling to neighboring SLs. Although theory predicts a sizable energy gap in the surface Dirac cone due to magnetism,
angle-resolved photoemission spectroscopy (ARPES) experiments rarely observe the surface gap \cite{otrokov2019prediction,PhysRevB.100.121104,PhysRevResearch.1.012011,shikin2021sample,chen2019intrinsic,PhysRevX.9.041038,PhysRevX.9.041039,PhysRevX.9.041040,PhysRevB.101.161109,PhysRevB.101.161113,Ji2021Detection}.
Despite the controversy on the surface magnetic gap, QAHE, Chern insulator, and AI were experimentally observed in MBT thin films \cite{deng2020science,liu2020robust,nwaa089,liu2021magnetic,ying2021experimental}.

\begin{figure}[tbp]
\includegraphics[width=0.7\columnwidth]{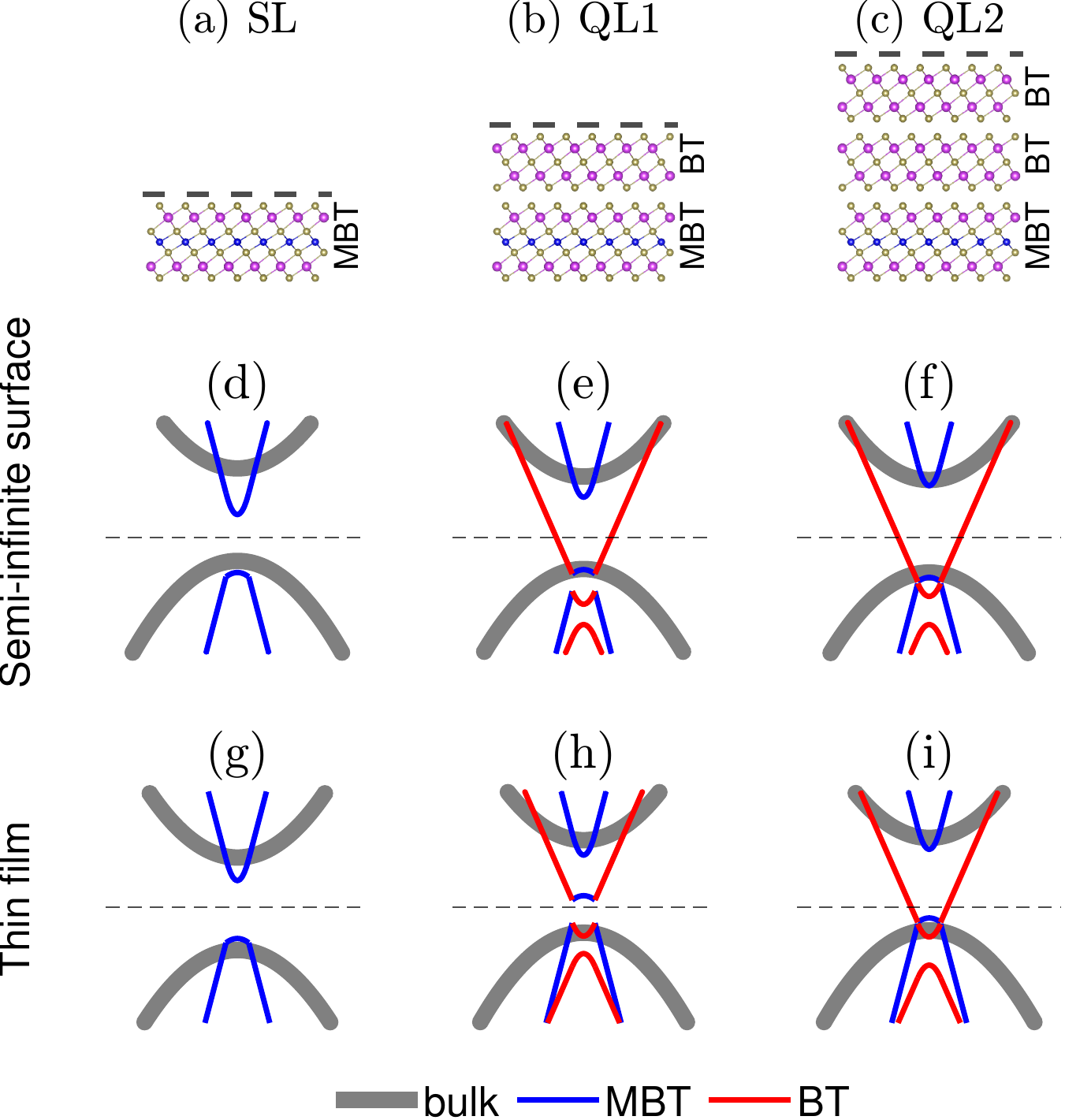}
\caption{\label{MBT_FIGURE1_SCHEMATICS} Surface crystal structures and schematic band structures. (a)-(c) show the surfaces with SL, QL1, and QL2 terminations, respectively. The black dashed line indicates the surface region above the bulk.
(d)-(f) are the schematic band structures of the corresponding surfaces in (a)-(c), where the grey, blue and red curves represent the bands of the bulk, surface MBT layer (the BT layer below the MBT can also be included) and BT layer(s) on top. The black dashed line indicates the Fermi energy. (g)-(i) are the schematic band structures of the corresponding thin films.
For the QL1 surface in (e), the surface hybridization between the surface BT and MBT layers leads to the surface hybridization gap, which is filled by the bulk valence band leading to a gapless electronic structure.
This hybridization gap is exposed in thin film case leading to a gapped electronic structure in (h).
For the QL2 surface and film, the hybridization between the surface BT and MBT layers is too weak to open a sizable hybridization gap, leading to a metallic band structure even for a thin film.
}
\end{figure}

More recently, a family of materials were synthesized by inserting Bi$_2$Te$_3$ layers (BT, or quintuple layer, QL) into MnBi$_2$Te$_4$ to form superlattice like crystals (MnBi$_2$Te$_4$)(Bi$_2$Te$_3$)$_n$ ($n=1,2,3$) and likely stabilize the interlayer FM order \cite{klimovskikh2020tunable,hu2020sciadv,wu2020toward}.
However, ARPES \cite{PhysRevB.101.161113,PhysRevX.9.041065,XU20202086,PhysRevX.10.031013,hu2020van,PhysRevB.102.045130,PhysRevLett.126.176403} still revealed gapless surface states on the MBT-terminated surface (labeled as the SL surface) for $n \ge 1$ compounds, similar to pure MBT, for which the surface magnetism is speculated to change or disappear.
Furthermore, whether BT-terminated surfaces are gapped or gapless is also under debate between different experiments.
For example, covering one BT layer on top of the SL surface forms the QL1 surface [see Fig. \ref{MBT_FIGURE1_SCHEMATICS}(b)], and covering two BT layers on top of the SL surface forms the QL2 surface [Fig. \ref{MBT_FIGURE1_SCHEMATICS}(c)].
So far, both experiments and calculations agree that the QL2 surface is gapless \cite{gordon2019strongly,klimovskikh2020tunable,PhysRevB.101.161113,PhysRevB.102.035144,PhysRevB.102.245136,hu2020sciadv,zhong2021light}. However, many ARPES experiments and calculations concluded that the QL1 surface is gapped \cite{hu2020van,klimovskikh2020tunable,PhysRevX.10.031013,PhysRevX.9.041065,gordon2019strongly,zhong2021light,hu2020sciadv} while some other works claimed the gapless feature for the same surface \cite{PhysRevB.101.161113,PhysRevX.9.041065,XU20202086,PhysRevB.102.035144,PhysRevB.102.045130,PhysRevB.102.245136}.
We note that the gapped nature of surface states is crucial to realizing QAHE, Chern insulator, or AI, which requires insulating thin films including their surfaces.

In this work, we focus on BT-terminated surfaces and find that they are actually gapless for (MnBi$_2$Te$_4$)(Bi$_2$Te$_3$)$_n$ ($n=1,2,3$) by calculations and propose a general scenario (see Fig. \ref{MBT_FIGURE1_SCHEMATICS}) to describe their surface band structure.
A mismatch between the surface hybridization gap and bulk energy gap removes the existence of a global gap on BT-terminated surfaces.
The existence of bulk states inside the surface hybridization gap rationalizes the origin of contradictions in previous experiments and theory. In theory, the insulating phase can still be realized in a film if both surfaces are MnBi$_2$Te$_4$ layers. Otherwise, it requires that the film thickness is less than 10$\sim$20 nm to increase the bulk gap.

%\section{Methods}

%\section{Results}
$Results.$
We first summarize the general scenario to describe surface band structures in theory. Here we use the abbreviations MBT and BT to represent respective van der Waals layers and use SL, QL1 and QL2 to represent related surfaces as shown in Fig. \ref{MBT_FIGURE1_SCHEMATICS}(a)-(c).
The surface electronic structures can be described by three sets of valence and conduction bands: the surface MBT bands, the surface BT bands, and bulk bands, as indicated in Fig. \ref{MBT_FIGURE1_SCHEMATICS}(d)-(f).

(i) For the SL surface [Fig. \ref{MBT_FIGURE1_SCHEMATICS}(a)], only two sets of the bands ($i.e.$ the bands of the surface MBT layer and bulk) are available, where according to our calculation, the conduction band minimum is contributed solely by the surface MBT layer and the valence band maximum is contributed by the bulk as shown in Fig. \ref{MBT_FIGURE1_SCHEMATICS}(d).

(ii) When one BT layer covers MBT and forms the QL1 surface [Fig. \ref{MBT_FIGURE1_SCHEMATICS}(b), $e.g.$, in MnBi$_4$Te$_7$], the surface BT bands arise.
Because the surface BT layer is close to the underlying MBT layer, their bands hybridize heavily, resulting in a sizable surface hybridization gap.
However, such a hybridization gap of the surface states on the QL1 surface is filled by bulk valence bands, removing the global band gap, as shown in Fig. \ref{MBT_FIGURE1_SCHEMATICS}(e).

\begin{figure}[tbp]
\includegraphics[width=1\columnwidth]{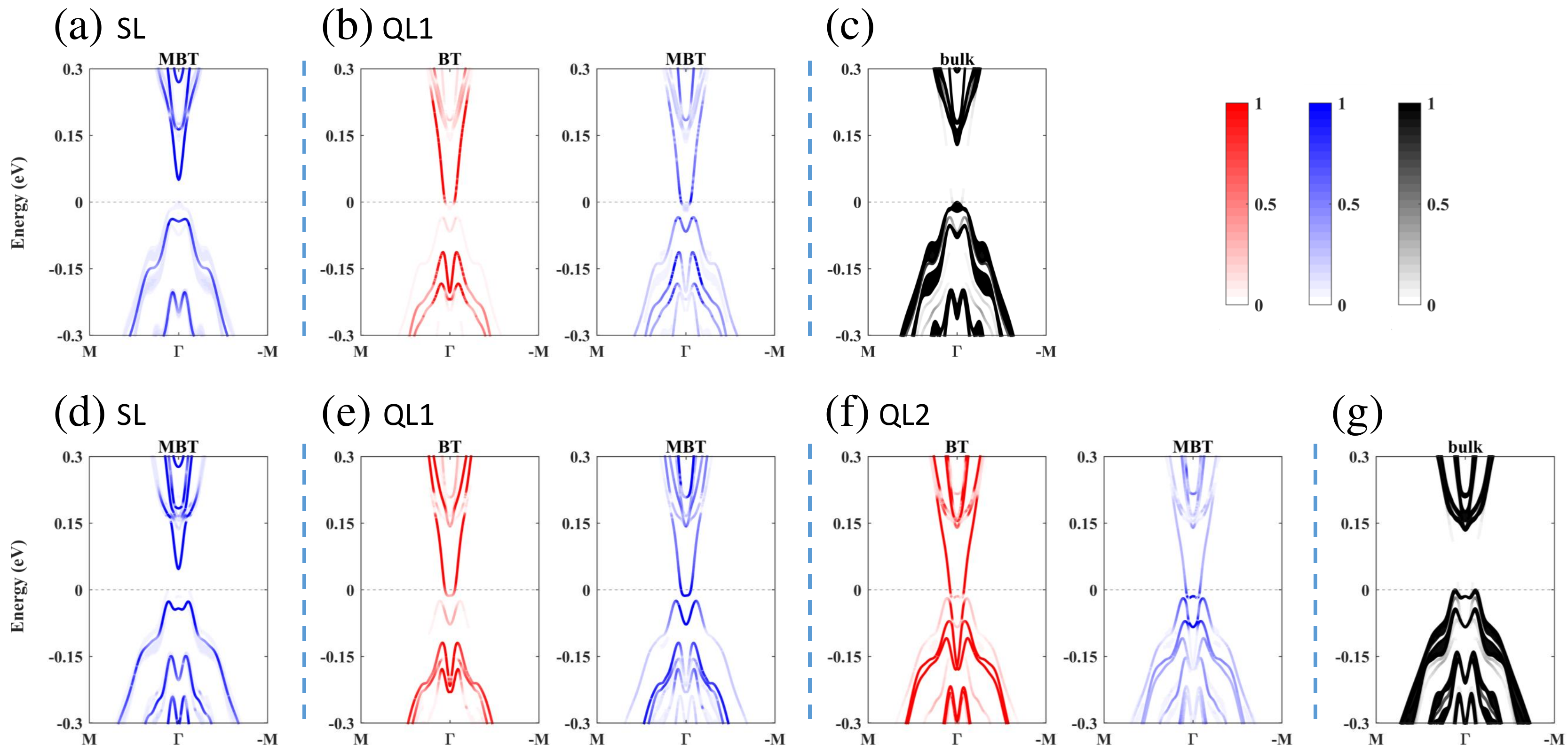}
\caption{\label{MBT_FIGURE2_SURFACE} Surface layer-specific contributions to the band structures of different thick ($\sim$ 60 nm) films as obtained from the tight-binding Hamiltonian calculations. (a)-(b) are for the SL and QL1 films of MnBi$_4$Te$_7$ respectively. The bulk contribution in (a) and (b) are the same, as shown in (c).
(d)-(f) are the surface layer-specific band structures for the SL, QL1, and QL2 films of MnBi$_6$Te$_{10}$ respectively. The bulk contributions in (d)-(f) are also the same, as shown in (g).
The surface layers and bulk are defined in the same way as those in Fig. \ref{MBT_FIGURE1_SCHEMATICS}.
The maximum of the bulk valence bands is set to energy zero as indicated by the black dashed line.
The color maps are shown in the upper right corner.
}
\end{figure}

(iii) The QL2 surface in Fig. \ref{MBT_FIGURE1_SCHEMATICS}(c) ($i.e.$ covering two BT layers on top of the SL surface, $e.g.$ in MnBi$_6$Te$_{10}$) is different from the case of QL1.
Due to the large distance between the surface BT layer and the lower MBT layer, the hybridization between them is too weak to open a large hybridization gap, which is also overwhelmed by bulk valence bands.
Thus, QL2 exhibits a metallic surface band structure even without an appreciable hybridization gap [see Fig. \ref{MBT_FIGURE1_SCHEMATICS}(f)], which is consistent with ARPES experiments. The case of covering three (or more) BT layers on top of the SL surface [$e.g.$ in MnBi$_8$Te$_{13}$ ($n = 3$)] is similar to the QL2 surface in Fig. \ref{MBT_FIGURE1_SCHEMATICS}(f), thus we do not discuss more on such a surface.

We now demonstrate the above scenario from first-principles calculations for different surfaces. Since we are interested in QL1 and QL2 surfaces, we will focus on MnBi$_4$Te$_{7}$ and MnBi$_6$Te$_{10}$ in the following. We construct tight-binding Hamiltonians from the respective bulk tight-binding Hamiltonians for super-thick (about 60 nm) films [see the methods in the Supplemental Material (SM) \cite{SM} for more information]. The film is terminated by the same type of layers on both surfaces. We label the film as the SL/QL1/QL2 film according to the surface termination.
The surface layer-specific band structures as that in Fig.\ref{MBT_FIGURE1_SCHEMATICS}(d)-(f) are obtained by projecting the film band structures onto the surface layers, where the remaining bands are regarded as the bulk contribution of the thick films.
The results are shown in Fig. \ref{MBT_FIGURE2_SURFACE}.

For SL surfaces, the global surface gap is merely 50 meV, although the MBT layer gap is about 80 meV [Fig. \ref{MBT_FIGURE2_SURFACE}(a) and (d)] and the bulk gap is around 130 meV [Fig. \ref{MBT_FIGURE2_SURFACE}(c) and (g)]. The reduction of the energy gap is induced by the mismatch between the MBT gap and the bulk gap.
One can find that the conduction band minimum is solely contributed by the surface MBT layer and the valence band maximum is contributed by the bulk.

For the QL1 surfaces, Fig. \ref{MBT_FIGURE2_SURFACE}(b) and (e) show a pronounced surface hybridization gap of about 28 meV and 11 meV for MnBi$_4$Te$_7$ and MnBi$_6$Te$_{10}$, respectively, which are induced by the hybridization between the surface BT and MBT layers.
Surface bands right above (below) the hybridization gap are mainly contributed by the surface BT (MBT) layer.
The hybridization between surface BT and MBT layers is further demonstrated by the prominent evolution of the surface band structures with gradually reducing the interaction between them.
In detail, the upper branches contributed by the BT layer cross the Fermi energy continuously when the interaction is small, which eliminates the hybridization gap (see the SM \cite{SM} for details).
Such a hybridization gap overlaps with the bulk valence bands near the $\Gamma$ point, resulting in a gapless surface.
We note that the bulk valence band maximum is right at $\Gamma$ point in MnBi$_4$Te$_7$ while it is slightly off $\Gamma$ point in MnBi$_6$Te$_{10}$. Because continual bulk bands exist far from the surface in space, surface states and bulk states cannot open a global hybridization gap even if they hybridize in the surface region.

For the QL2 surface in MnBi$_6$Te$_{10}$, the hybridization between the two BT layers on the surface and the MBT layer beneath is weak, leading to a small surface hybridization gap \cite{SM}.
In addition, MBT bands almost merge into bulk bands.
Such a weak hybridization is also confirmed by the non-prominent changes of the surface band structures with the decrease of the interaction between the surface layers \cite{SM}.
This can be explained by the large distance ($\sim$22 \AA) between the MBT layer and the surface BT layer.

\begin{figure}[tbp]
\includegraphics[width=1\columnwidth]{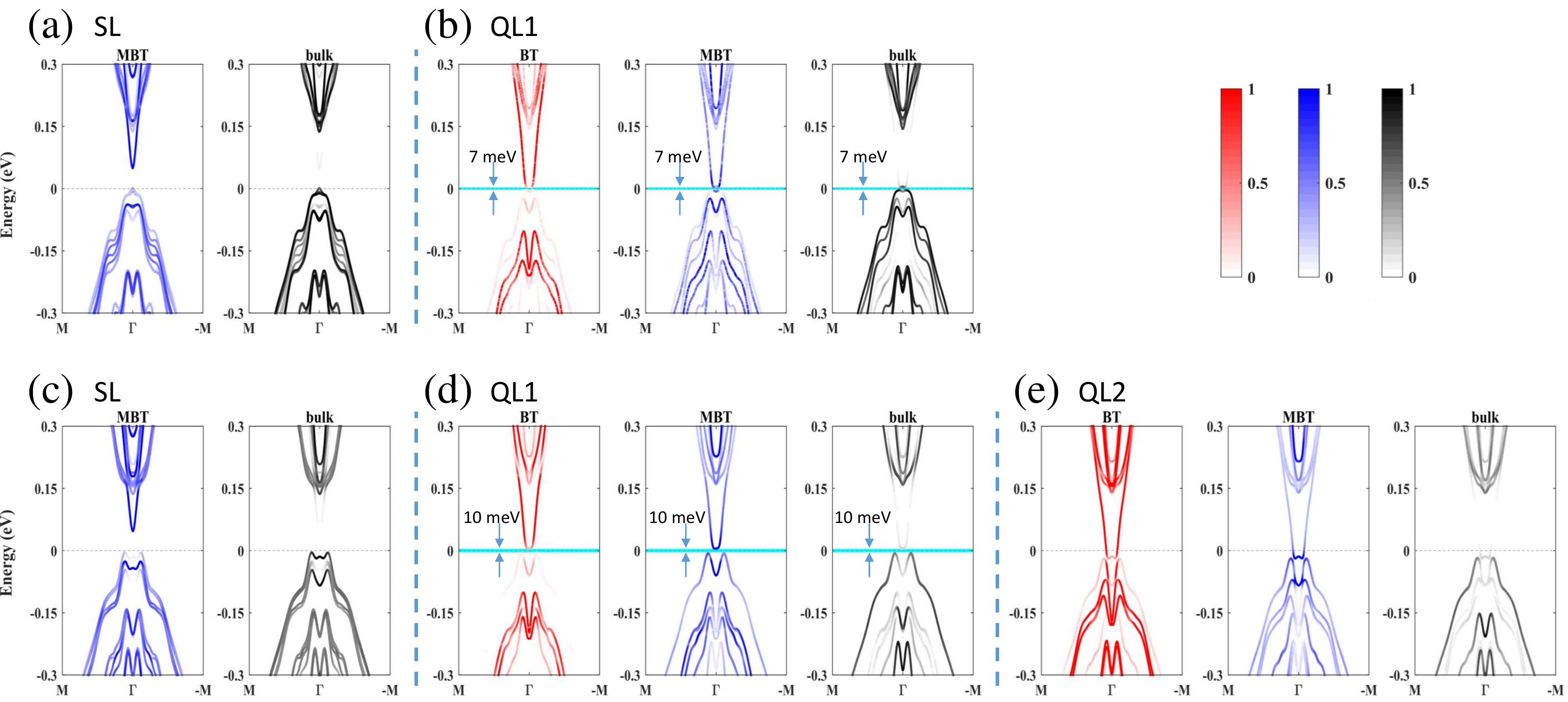}
\caption{\label{MBT_FIGURE3_VASP_film} Similar to Fig. \ref{MBT_FIGURE2_SURFACE} but for thin films, $i.e.$ (a)-(b) for the SL and QL1 films of MnBi$_4$Te$_7$ and (c)-(e) for the SL, QL1 and QL2 films of MnBi$_6$Te$_{10}$. For QL1 surface in (b) and (d), the small band gap is indicated by the cyan area (7 meV for MnBi$_4$Te$_7$ and 10 meV for MnBi$_6$Te$_{10}$). Notice that the bulk contributions of the different films of the same material are different.
}
\end{figure}

Figure \ref{MBT_FIGURE2_SURFACE} confirms the band alignment scenario, although real band structures are complicated.
The gapless nature of QL surfaces poses challenges to realizing QAHE/AI in $n \ge 1$ compounds unless both surfaces can be selected to SLs. However, there is still a possibility to obtain a gapped QL surface. The gapless feature is induced by bulk bands overlapping the surface hybridization gap on the QL surface.
If there is a way to remove bulk valence bands from the hybridization gap, then a gapped QL surface can be obtained, as shown schematically in Fig. \ref{MBT_FIGURE1_SCHEMATICS}(h) for the QL1 surface.
As we will show below, the size effect can play such a role.
However, pushing down bulk bands cannot lead to a gap for the QL2 surface due to the diminishing hybridization gap [see Fig. \ref{MBT_FIGURE1_SCHEMATICS}(i)]. We may introduce surface-surface interaction to open the band gap on the QL2 surface which is technically achievable by the reduction of the film thickness.
Unfortunately, our calculations show that for the thinnest QL2 film, no band gap is opened \cite{SM}.

Now we show the gap opening due to the size effect in thin films.
We still consider the three films above but with much smaller thicknesses.
In Figure~\ref{MBT_FIGURE3_VASP_film}, the thicknesses of the SL and QL1 films of MnBi$_4$Te$_7$ are about 105 and 102 \AA~, respectively, and the thicknesses of the SL, QL1, and QL2 films of MnBi$_6$Te$_{10}$ are about 78, 65, and 85 \AA~, respectively.
The surface layer-specific band structures and bulk contributions of these thin films as shown in Fig. \ref{MBT_FIGURE3_VASP_film} are obtained in a similar way as that in Fig. \ref{MBT_FIGURE2_SURFACE}.
For the thin SL films in Fig. \ref{MBT_FIGURE3_VASP_film}(a) and (c), while the conduction band minimum is still contributed solely by the surface MBT layer, the contribution of the surface MBT layer to the valence band maximum is increased.
The lowering of the bulk valence bands is shown in the QL1 films in Fig. \ref{MBT_FIGURE3_VASP_film}(b) and (d), where the surface hybridization gap is partly released by the bulk valence bands, leading to a band gap of about 7 meV in Fig. \ref{MBT_FIGURE3_VASP_film}(b) and 10 meV in (d).
For the QL2 film in Fig. \ref{MBT_FIGURE3_VASP_film}(e), the bulk contribution to the valence band maximum is also decreased.
However, no sizable band gap appears due to the tiny surface hybridization gap.
Within such results, the most significant feature is the opening of the global band gap in thin QL1 films as compared to thick QL1 films.
This indicates that there is a critical thickness for the QL1 film, below which the QL1 film has a band gap while above which the film is gapless.

Up to now, we have rationalized the general scenario in Fig. \ref{MBT_FIGURE1_SCHEMATICS}, where the key features are that i) the hybridization between the surface layers determines the surface hybridization gap and ii) the alignment of the surface bands with the bulk bands determines the global band gap.
This general scenario provides a competitive explanation for the recent experimental controversy on QL1 surfaces.
In other words, whether the experimentally observed surface states of the QL1 surface are gapped or gapless may depend on whether the bulk states are detected.
ARPES with larger photon energy may be more accessible for bulk bands.
For example, recent ARPES experiments with photon energy of 6.3 $\sim$ 6.4 eV \cite{PhysRevX.10.031013,PhysRevLett.126.176403,klimovskikh2020tunable} detected a band gap for the QL1 surface.
With increasing the photon energy, for example, to 21 eV in Ref. \onlinecite{PhysRevB.102.035144} and $\sim$80 eV in Refs. \onlinecite{PhysRevLett.126.176403,PhysRevX.9.041065}, more states appear inside the surface hybridization gap, which can be interpreted as bulk states according to our theory.
We also notice that most previous calculations based on slab models \cite{gordon2019strongly,Sitnicka_2021,zhong2021light,PhysRevB.102.245136,PhysRevX.10.031013,hu2020sciadv,wu2019natural} found gapped surface states for the QL1 surface.
These calculations adopted slab models with thicknesses from 5 to 13 nm.
The size effect is still significant in this thickness regime. As we will show, QL1 slab models exhibit a gapless surface above 10$\sim$20 nm. A thinner slab may be useful to describe the band structure of a thin film, but it can be sometime misleading to understand surface bands on a semi-infinite surface.
For the other QL surfaces such as QL2 [Fig. \ref{MBT_FIGURE1_SCHEMATICS}(c)], the ARPES experiments consistently observed gapless surface states due to the absence of a pronounced surface hybridization gap.

\begin{figure}[tbp]
\includegraphics[width=0.9\columnwidth]{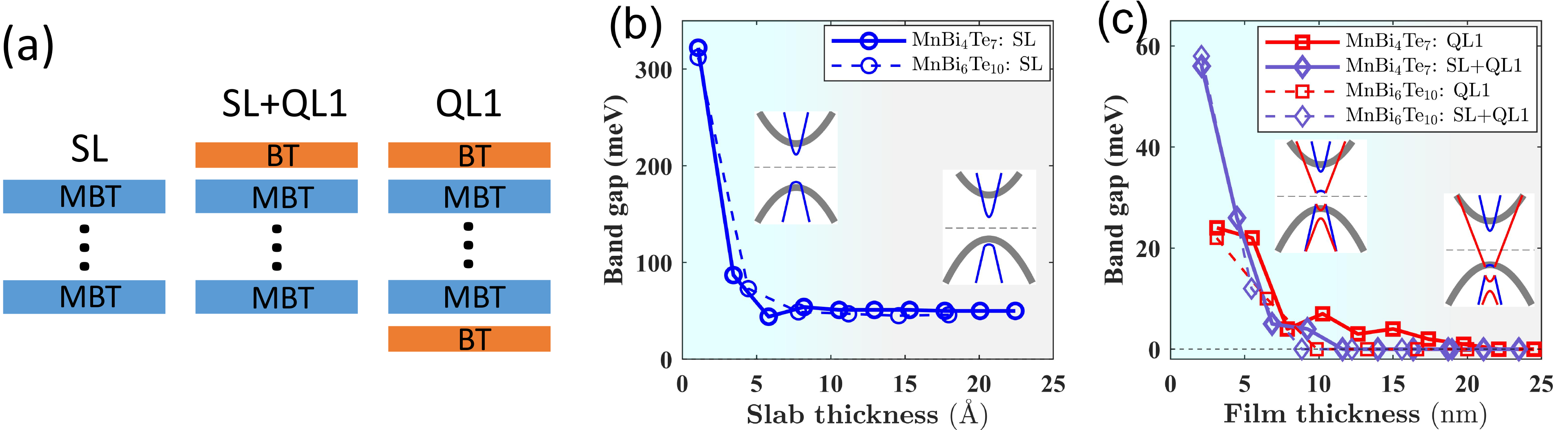}
\caption{\label{MBT_FIGURE4_QAH}
Schematic crystal structures and band gaps of the three films.
(a) shows the schematic structures of the three films, $i.e.$ SL (QL1) film with both surfaces being the SL (QL1) surface, and mix-surface film [SL+QL1] with the two surfaces being SL and QL1 respectively. Only the layers on the two surfaces are shown for simplicity.
(b) shows the band gap of the SL film as a function of the film thickness. The insets [$i.e.$ Fig. \ref{MBT_FIGURE1_SCHEMATICS}(d) and (g)] show the schematic band structures of the thick (light grey area) and thin (light cyan area) films.
(c) is similar to (b) but for QL1 and SL+QL1 films. The insets in (c) [Fig. \ref{MBT_FIGURE1_SCHEMATICS}(e) and (h)] show the schematic band structures of thick (light grey) and thin (light cyan) QL1 films.
}
\end{figure}

Our results not only rationalize recent ARPES experiments but also shed light on the realization of QAHE and AI in thin films of these materials, where a band gap is required.
Figure \ref{MBT_FIGURE1_SCHEMATICS} indicates that the SL surface should be always gapped with no film thickness limit, and the thin QL1 film has a band gap while the thick QL1 film or semi-infinite surface is gapless.
Thus, we propose three kinds of films that may have global band gaps, SL film, QL1 film, and the mix-surface film [SL+QL1] where one surface is SL and the other surface is QL1, as shown in Fig. \ref{MBT_FIGURE4_QAH}(a).
We calculated the band gap as a function of the film thickness and the results are shown in Fig. \ref{MBT_FIGURE4_QAH}(b) and (c).
We also employed full density functional theory (DFT) calculations based on slab models. While DFT produces similar results with tight-binding calculations for thin films, DFT calculations are very computationally demanding for the thick films (see the SM for details \cite{SM}) which are beyond our compatibility for the large number of thick films. Thus we employ the results from tight-binding calculations for consistency and completeness in Fig. \ref{MBT_FIGURE4_QAH}.

Figure \ref{MBT_FIGURE4_QAH}(b) and (c) show that the band gap decreases with increasing the film thickness, which agrees with the expectation that the bulk valence bands go up with increasing the film thickness.
In detail, the band gap of the SL film converges to $\sim$50 meV when the film thickness goes to the large limit.
For the thin SL films the valence band maximum is mainly contributed by the surface MBT layers and for thick films the bulk constitutes the valence band maximum, as indicated by the insets in Fig. \ref{MBT_FIGURE4_QAH}(b).
The band gaps of the QL1 film and mix-surface [SL+QL1] film converge to zero with the film thickness going to a large limit, indicating a critical thickness for such two films beyond which the films are gapless.
According to Fig. \ref{MBT_FIGURE4_QAH}(c), we estimate the critical thickness in the range of 10 to 20 nm for MnBi$_4$Te$_7$ and $\sim$ 10 nm for MnBi$_6$Te$_{10}$.

We emphasize that the critical thickness of the QL1 film could be slightly larger than that of the mix-surface film [SL+QL1] since for a thin [SL+QL1] film the SL surface may constitute the valence band maximum (higher than the bulk valence bands).
Here we ignore such differences in our estimation.
We also point out that in real material the critical thickness of the two films may be larger than the estimated values in this work, mainly caused by two reasons.
The first one is a physical reason that the surface charge relaxation in the real films cannot be considered in our tight-binding calculations.
The second one is a technical reason that the generalized gradient approximation used for the exchange-correlation functional to extract the tight-binding Hamiltonian generally underestimates the band gap.
Nevertheless, we believe that the general physical picture is valid.

Such a critical thickness provides guidance for experiments on realizing insulating topological phases.
In the sense of surface gap, MnBi$_2$Te$_4$ is the most promising one, since its thin film is naturally MBT terminated.
For (MnBi$_2$Te$_4$)(Bi$_2$Te$_3$)$_n$ with $n \ge 1$, the film requires one of the following conditions, (i) the two surfaces of the film are both SL, (ii) the thickness is below some critical value if the film has at least one QL1 surface but no QL2 termination.

%\section{Conclusions}
$Conclusions.$
In summary, we have addressed the band gap problem of the Bi$_2$Te$_3$ surfaces in (MnBi$_2$Te$_4$)(Bi$_2$Te$_3$)$_n$ with a general scenario.
We identified the essential role of the surface hybridization gap as well as its alignment with bulk bands in determining the global band gap, which is crucial for realizing the QAHE or AI. Such an alignment can be tuned by the size effect to modify the bulk gap. For such a film with Bi$_2$Te$_3$ surfaces, the critical thickness of the size effect is estimated to be in the range of 10 to 20 nm. No sizable band gap is obtained for the MnBi$_2$Te$_4$ surface covered by more than one Bi$_2$Te$_3$ layer due to the absence of an apparent surface hybridization gap. Our work provides practical design principles for topological devices.

%\section{acknowledgment}
\textbf{Acknowledgment.}
We thank kind help from Daniel Kaplan and fruitful discussions with Shuolong Yang, Yong Xu, and Z.X. Shen. B.Y. acknowledges the financial support by the European Research Council (ERC Consolidator Grant ``NonlinearTopo'', No. 815869), and ISF - Singapore-Israel Research Grant (No. 3520/20).

%\bibliography{Reference_MBT}
%merlin.mbs apsrev4-1.bst 2010-07-25 4.21a (PWD, AO, DPC) hacked
%Control: key (0)
%Control: author (0) dotless jnrlst
%Control: editor formatted (1) identically to author
%Control: production of article title (0) allowed
%Control: page (1) range
%Control: year (0) verbatim
%Control: production of eprint (0) enabled
%

\end{document}

\title{Supplemental Material for ``Facet dependent surface energy gap on magnetic topological insulators''}

\author{Hengxin Tan}
\author{Binghai Yan}
\affiliation{Department of Condensed Matter Physics, Weizmann Institute of Science, Rehovot 7610001, Israel}

%\date{\today}

\maketitle

\hspace*{\fill} \\
$Method.$ The electronic structures of the bulk materials are calculated with density functional theory (DFT) as implemented in the Vienna $ab$-$initio$ Simulation Package (\textsc{vasp}) \cite{VASP,PRB54p11169}, where the projected augmented wave \cite{PRB50p17953} potentials with 7 valence electrons for Mn, 5 valence electrons for Bi, and 6 valence electrons for Te are employed.
The exchange-correlation interaction between electrons is mimicked with the generalized gradient approximation as parameterized by Perdew-Burke-Ernzerhof \cite{PRL77p3865}.
A cutoff energy of 350 eV is used for the plane-wave basis set.
A Hubbard $U$ of 5 eV is used for the $d$ electrons of the Mn atoms.
The $k$-meshes of the bulk Brillouin zone sampling are 12$\times$12$\times$12 for MnBi$_2$Te$_{4}$, 12$\times$12$\times$4 for MnBi$_4$Te$_{7}$ ($n=1$), 9$\times$9$\times$9 for MnBi$_6$Te$_{10}$ ($n=2$), and 9$\times$9$\times$9 for MnBi$_8$Te$_{13}$ ($n=3$).
The experimental A-type AFM configuration is employed for (MnBi$_2$Te$_4$)(Bi$_2$Te$_3$)$_n$ with $n=0,1,2$ and the experimental FM configuration is employed for MnBi$_8$Te$_{13}$.
The zero damping DFT-D3 van der Waals correction \cite{DFTD3} is employed throughout the calculations.
Spin-orbital coupling is considered in all the electronic structure calculations except for structural relaxation.
The wannier function-based tight-binding Hamiltonians of the bulk materials are obtained by the maximally localized Wannier functions method as implemented in Wannier90 software \cite{wannier90} which is interfaced to \textsc{vasp}. The electronic structures of the films are calculated with the slab tight-binding Hamiltonians as constructed from the corresponding bulk tight-binding Hamiltonian.

\begin{figure}
\centering
\includegraphics[width=0.95\linewidth]{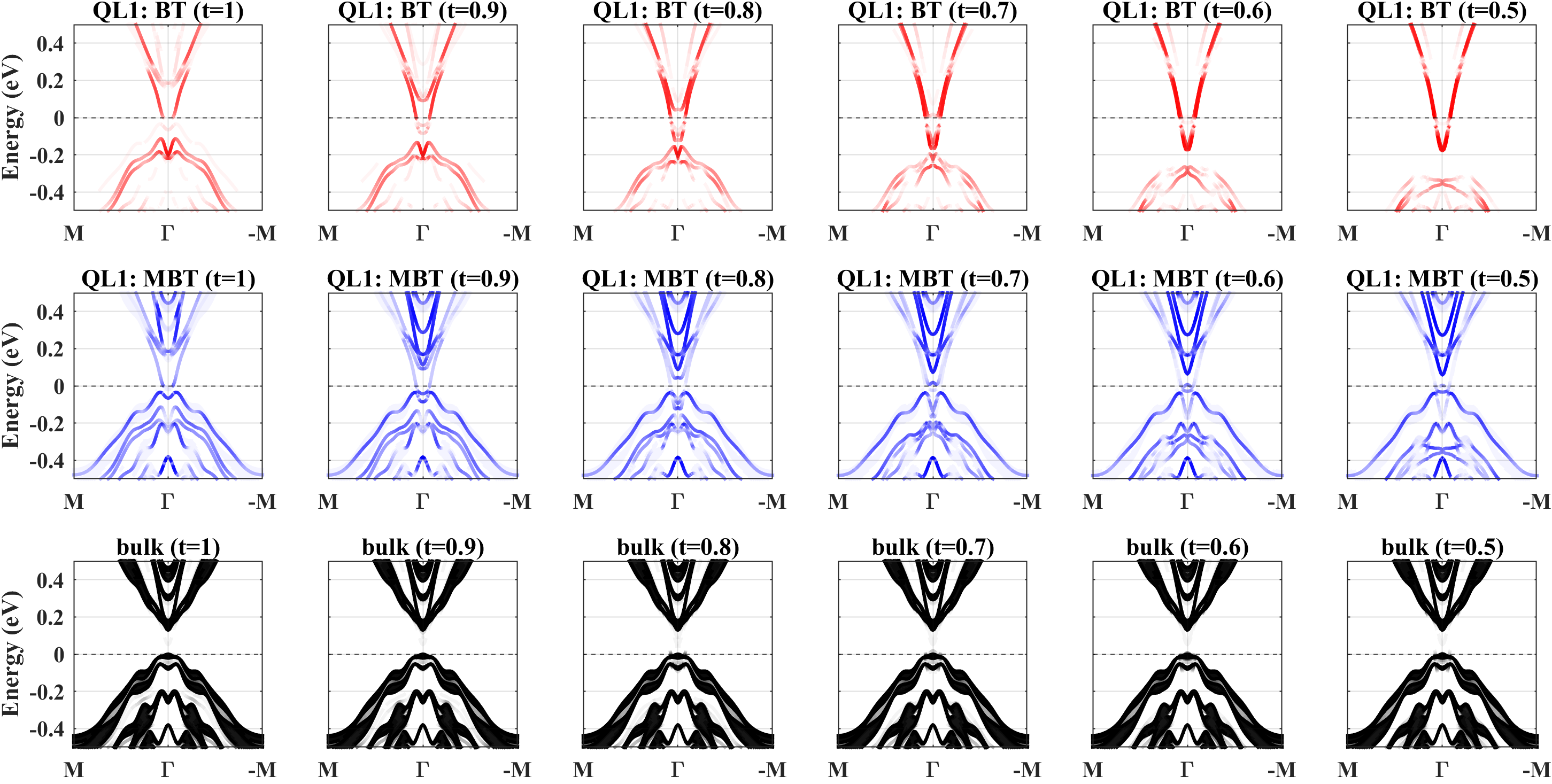}
\caption{\label{SMFIG-147-QL1}
Evolution of the surface layer-specific band structure of the QL1 surface of MnBi$_4$Te$_7$ with respect to the normalized interaction ($t$) between the surface BT layer and the MBT layer beneath. The surface BT, MBT layers, and bulk region of the film are defined in the same way as in the main text. The interaction $t$ of 1 corresponds to the normal interaction without any changes while $t=0$ means zeros interaction between the surface BT layer and the rest of the film. The band structures for $t~\textless~0.5$ are not shown since those band structures near the Fermi energy are similar to that in $t=0.5$. The prominent evolution of the band structures of the surface BT and MBT layers with respect to their interaction indicates the heavy hybridization between them.
}
\end{figure}

\begin{figure}
\centering
\includegraphics[width=0.95\linewidth]{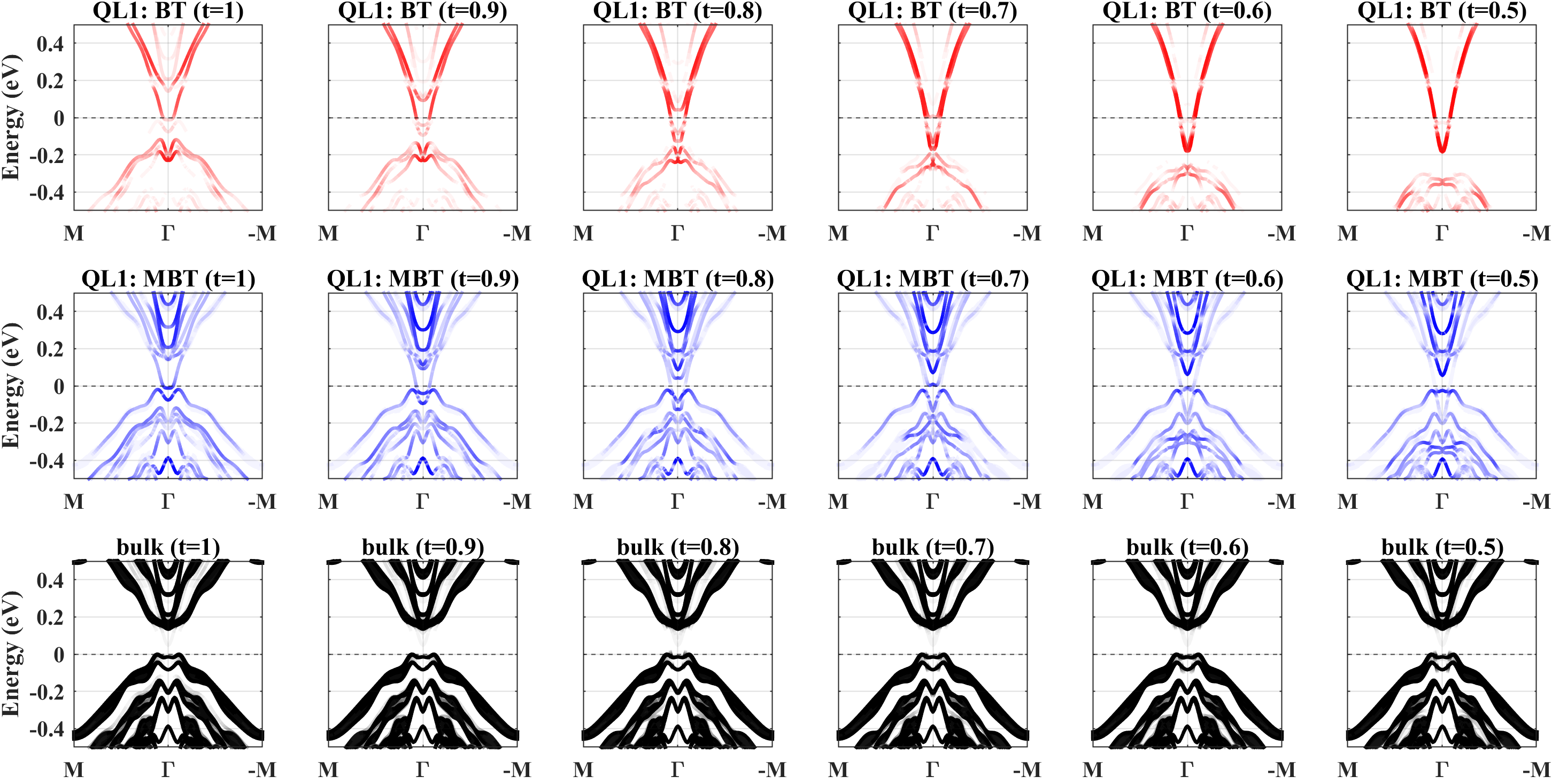}
\caption{\label{SMFIG-1610-QL1}
Similar to Fig. \ref{SMFIG-147-QL1} but for the QL1 surface of MnBi$_6$Te$_{10}$, which also indicates the heavy hybridization between the surface BT layer and the MBT layer beneath.
}
\end{figure}

\begin{figure}
\centering
\includegraphics[width=0.95\linewidth]{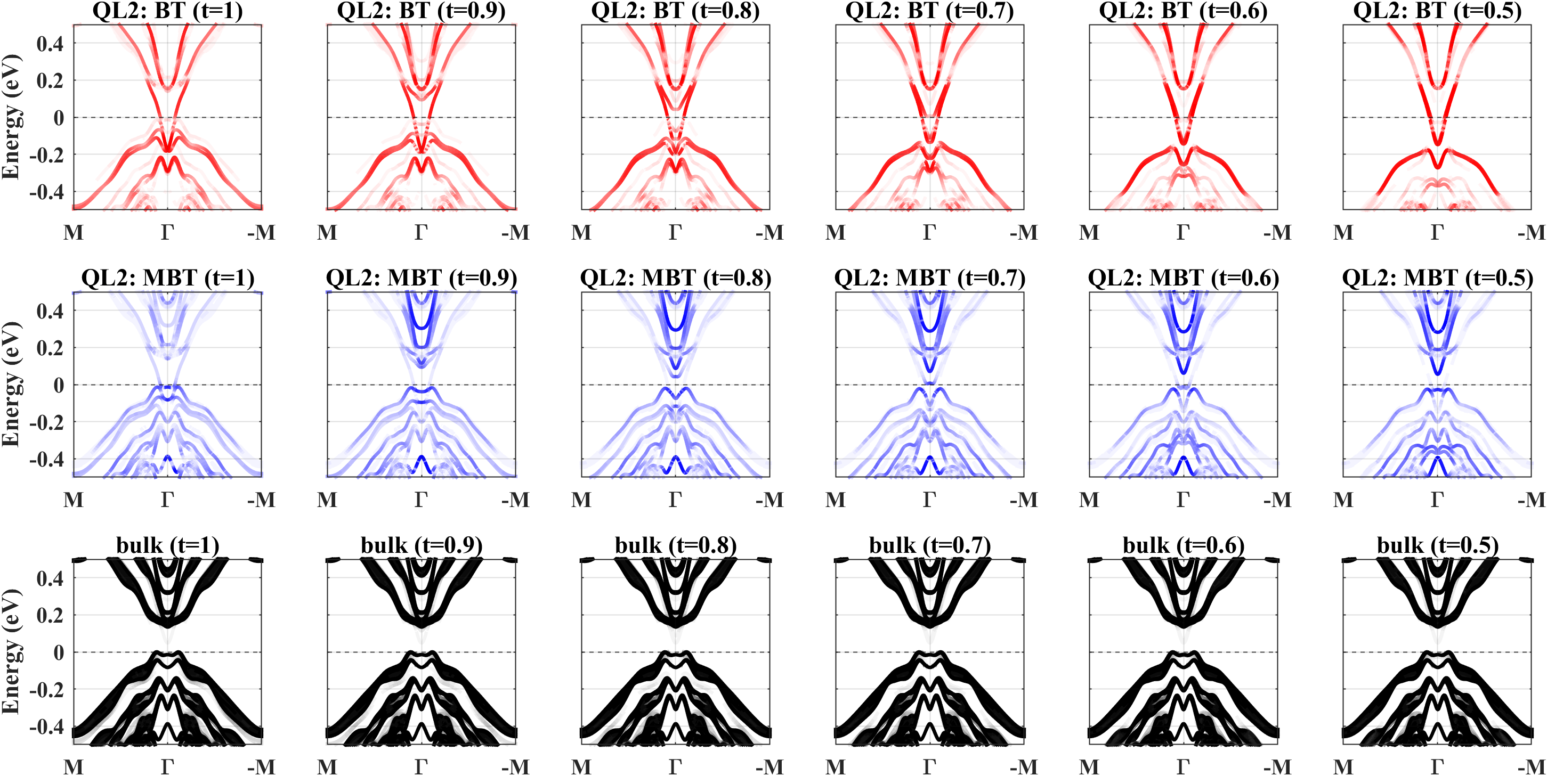}
\caption{\label{SMFIG-1610-QL2}
Similar to Fig. \ref{SMFIG-147-QL1} but for the QL2 surface of MnBi$_6$Te$_{10}$. As we can see that the band evolution with respect to the interaction between the two surface BT layers and the rest of the film is less prominent, especially for the two surface BT layers (the top row). Thus we deduce that the hybridization between the two surface BT layers and the MBT layer beneath is relatively weak, which is responsible for the diminishing hybridization band gap of the QL2 surface.
}
\end{figure}

\begin{figure}
\centering
\includegraphics[width=0.95\linewidth]{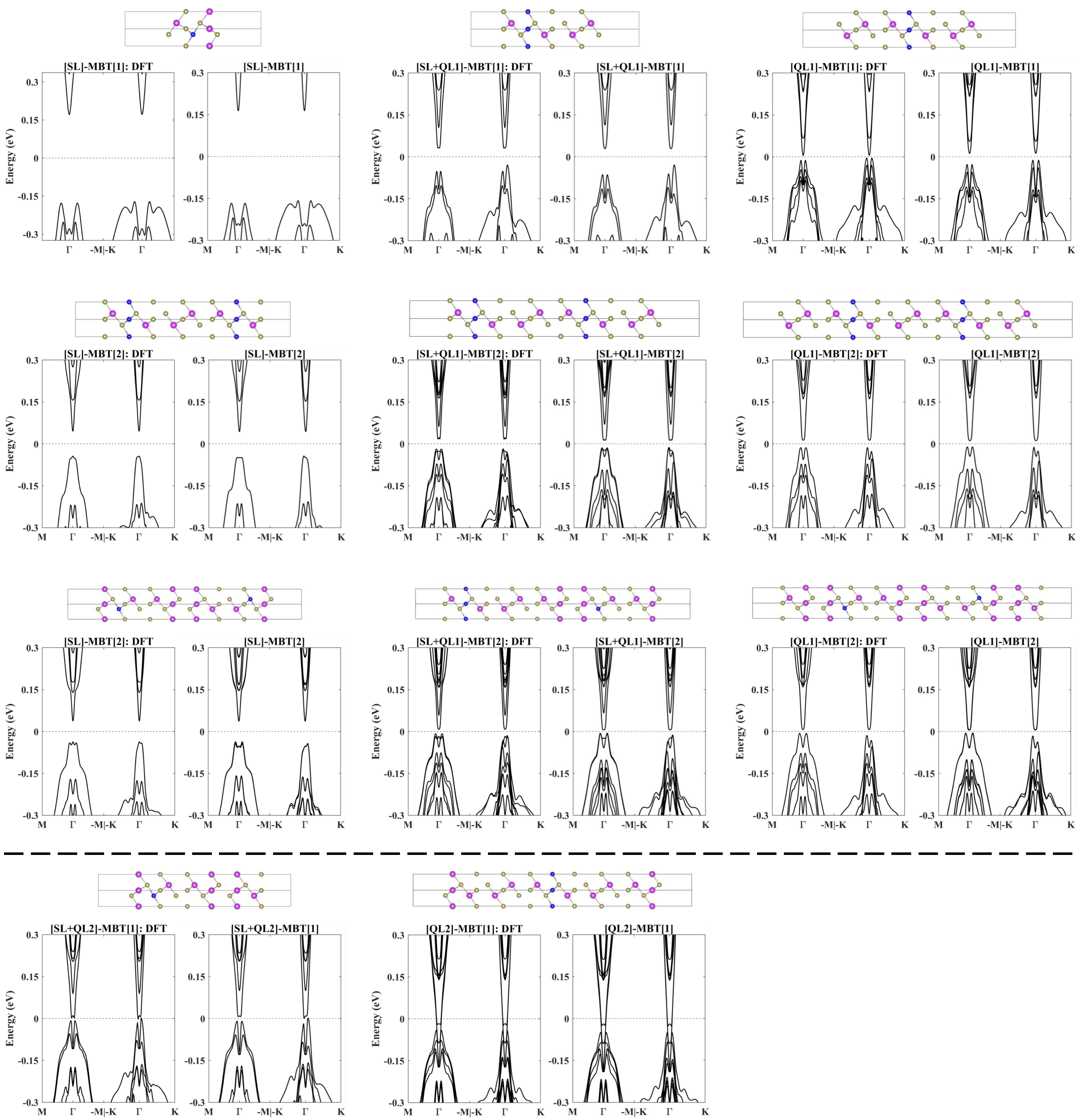}
\caption{\label{SMFIG-VP-TB}
Band structures of thin films. In each panel, the top part shows the film structure, the bottom left shows the band structure from density functional theory (DFT) calculation while the bottom right part shows the band structure from slab tight-binding Hamiltonian calculation. For films with a band gap, the middle of the band gap is set as the energy zero point. While the first row is available for all systems, the second row is unique to MnBi$_4$Te$_7$ and the third row is unique to MnBi$_6$Te$_{10}$. The band structures of the thinnest films with at least one QL2 surface are shown in the fourth row. The good agreement between the DFT band structures and the tight-binding band structures indicates the predictive power of our tight-binding Hamiltonian calculations for films.
The title of each band structure indicates the type of the film as defined in the main text as well as the number of MBT layers.
Notice that the SL and QL1 and QL2 films with one MBT layer have inversion symmetry while SL and QL1 films with 2 MBT layers have the combination of the inversion symmetry and time-reversal symmetry.
}
\end{figure}

\begin{figure}
\centering
\includegraphics[width=0.95\linewidth]{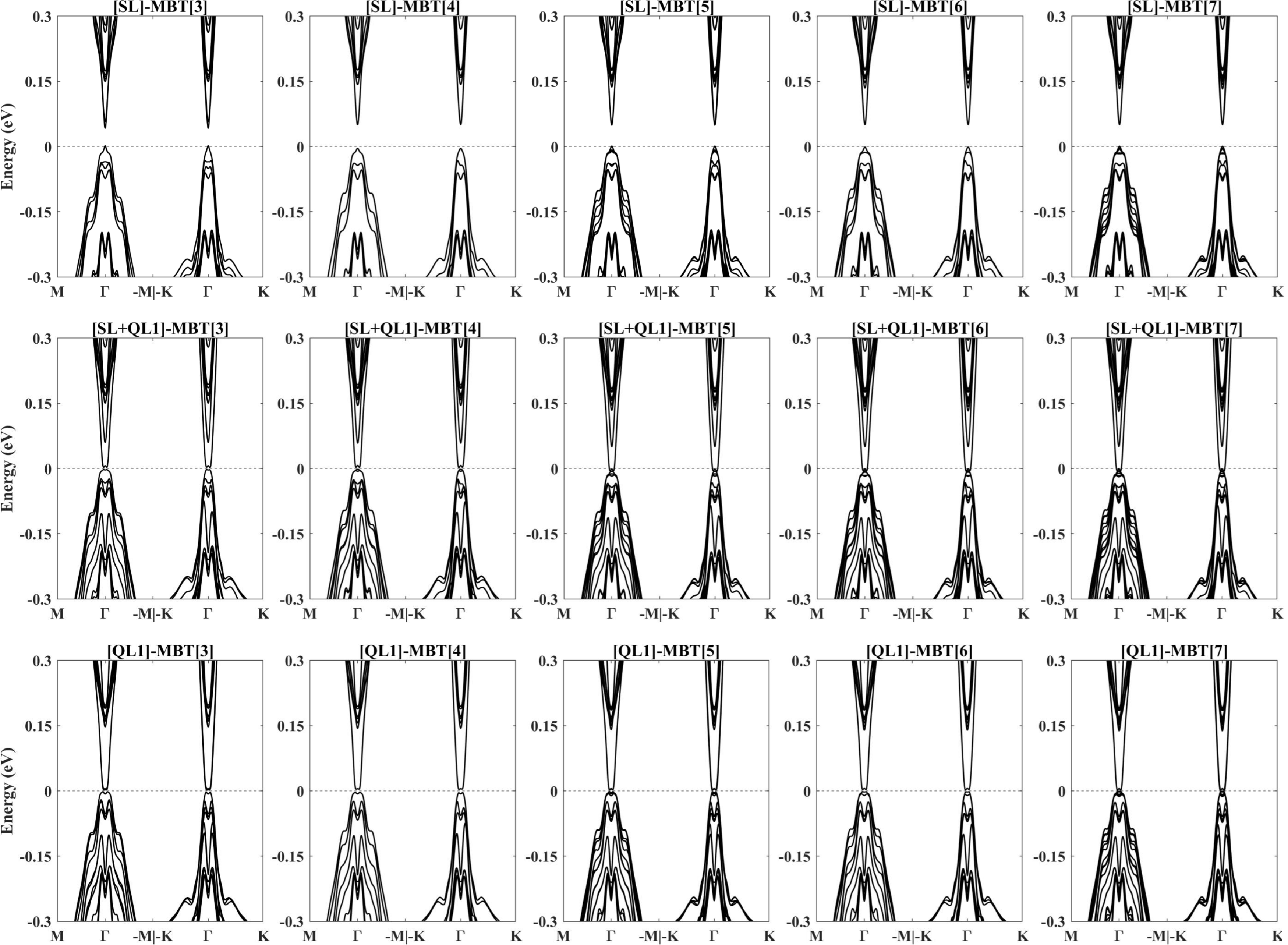}
\caption{\label{SMFIG-147-TB}
Band structures of some relatively thick films of MnBi$_4$Te$_7$ as calculated by slab tight-binding Hamiltonian. The valence band maximum is set as the energy zero for the SL films while the Fermi energy is set as energy zero for films without a band gap. The title of each band structure indicates the type of the film as defined in the main text as well as the number of MBT layers, which can be used to reproduce the films. Notice that the thin QL1 and SL+QL1 films may have very small band gaps. Also, notice that the SL and QL1 films with an odd number of MBT layers have inversion symmetry while SL and QL1 films with an even number of MBT layers have the combination of the inversion symmetry and time-reversal symmetry.
}
\end{figure}

\begin{figure}
\centering
\includegraphics[width=0.95\linewidth]{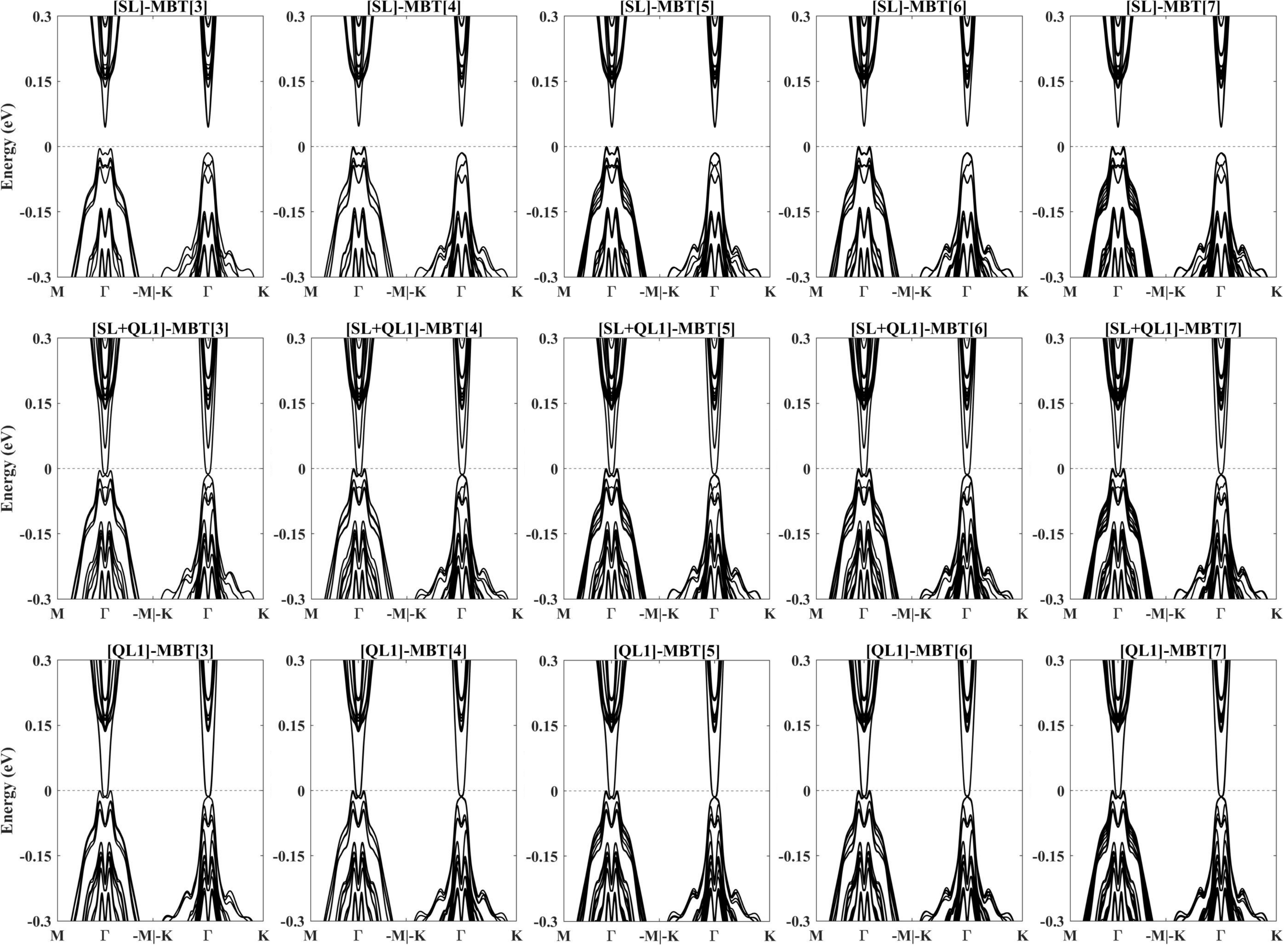}
\caption{\label{SMFIG-1610-TB}
Similar to Fig. \ref{SMFIG-147-TB} but for MnBi$_6$Te$_{10}$.
}
\end{figure}

\begin{figure}
\centering
\includegraphics[width=0.95\linewidth]{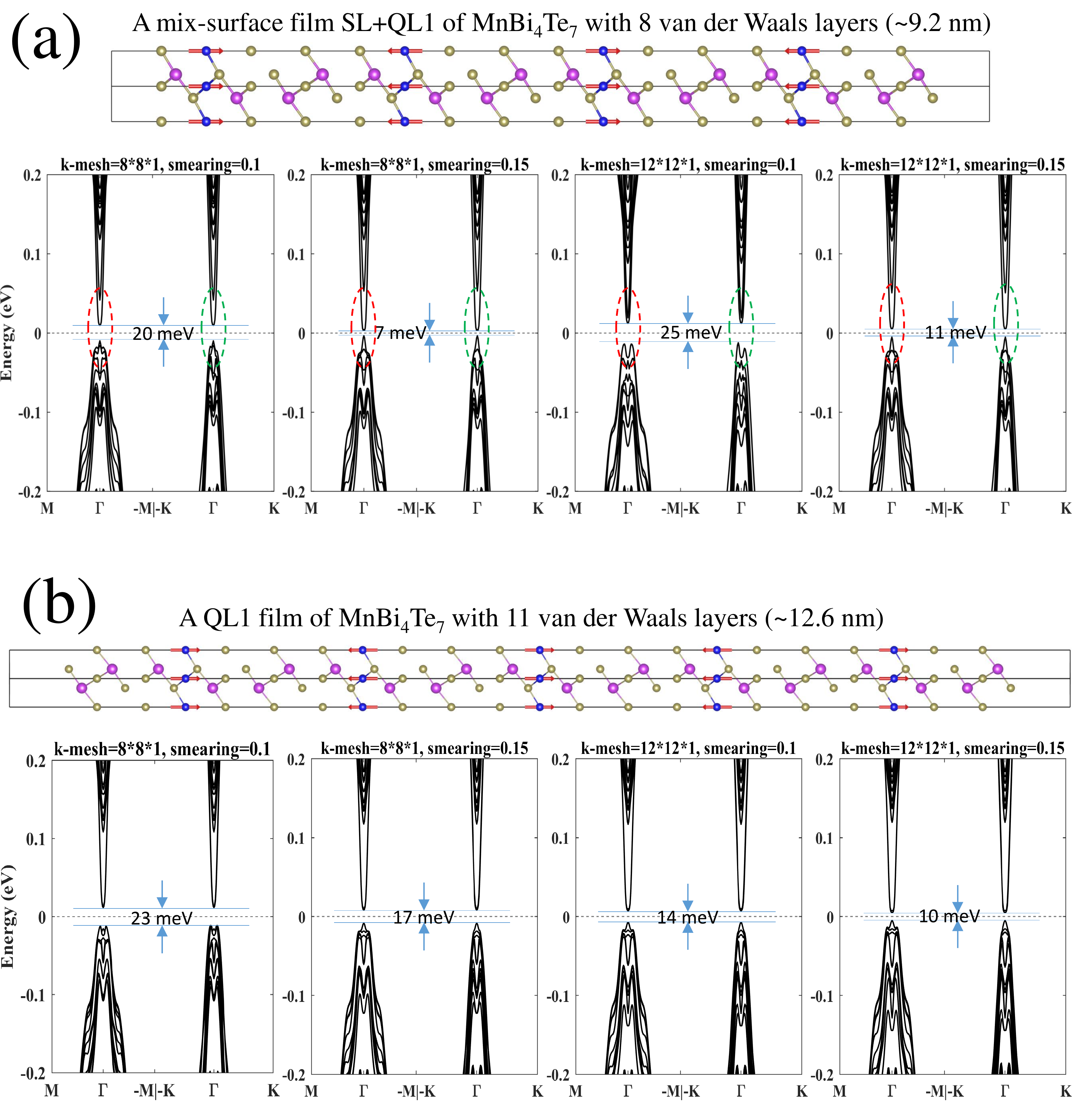}
\caption{\label{SMFIG-VP-thick-slab}
Two test cases of films of MnBi$_4$Te$_7$ as calculated by density functional theory as implemented in VASP with different $k$-meshes and smearing values for the electron distribution function.
(a) for a mix-surface film SL+QL1 with 8 van der Waals layers ($\sim$9.2 nm) which breaks both inversion and time-reversal symmetries nor has it their combination. The upper panel shows the structure of the film whereas the red vectors show the AFM configuration used in calculations. The four lower panels are the band structures under different $k$-meshes and smearing values as indicated in the title of each subplot. Notice that not only is the band gap (indicated in each subplot) sensitive to the parameters, but also the band edges ($e.g.$ the position of the valence band maximum) are changed prominently which are actually caused by the changes of the relative positions (in energy space) of the band structures of the two surfaces and bulk.
(b) is similar to (a) but for a QL1 film with 11 van der Waals layers ($\sim$12.6 nm) which has inversion symmetry. It can be seen that the band gap is sensitive to the calculating parameters.
Notice that the band gaps from the tight-binding calculations are 4 meV for (a) and 3 meV for (b).
We deduce that the sensitivity of the band gaps and band edges of the thick films is caused by the fact that the low-energy band structures of a thick film near the Fermi energy are very localized around the Brillouin zone center, which demands very high accuracy of the parameters to simulate accurately the charge density.
Such a large number of highly accurate density functional theory calculations for thick films are beyond our current computing power.
Thus the tight-binding results are employed in this work for the consistency and completeness of the data, which captures the right physics behind it as well.
}
\end{figure}

%%%%%%%%%%%%%%%%%%%%%%%%%%%%%%%%%%%%%%%%%%%%%%%%%%%%%%%%%%%%%%%%%%%%%%%%%%%%%%%%%%%%%%%%
\hspace*{\fill} \\
\hspace*{\fill} \\
\textbf{Contents for figures:}
\begin{itemize}
  \item Figure \ref{SMFIG-147-QL1} shows the evolution of the layer-specific band structures with respect to the interaction between the surface BT layer and the MBT layer  of the QL1 surface of MnBi$_4$Te$_7$.
  \item Figure \ref{SMFIG-1610-QL1} is similar to Fig. \ref{SMFIG-147-QL1} but for the QL1 surface of MnBi$_6$Te$_{10}$.
  \item Figure \ref{SMFIG-1610-QL2} is also similar to Fig. \ref{SMFIG-147-QL1} but for the QL2 surface of MnBi$_6$Te$_{10}$.
  \item Figure \ref{SMFIG-VP-TB} shows the band structures of thin films as obtained from both density functional theory (DFT) and tight-binding Hamiltonian calculations.
  \item Figure \ref{SMFIG-147-TB} shows the band structures of relatively thick films of MnBi$_4$Te$_7$ as obtained from tight-binding Hamiltonian calculations.
  \item Figure \ref{SMFIG-1610-TB} is similar to Fig. \ref{SMFIG-147-TB} but for MnBi$_6$Te$_{10}$.
  \item Figure \ref{SMFIG-VP-thick-slab} shows the band structures of two test cases of thick films of MnBi$_4$Te$_7$, which shows that the density functional theory calculations for thick films with at least one QL1 surface is computationally demanding.
\end{itemize}
%%%%%%%%%%%%%%%%%%%%%%%%%%%%%%%%%%%%%%%%%%%%%%%%%%%%%%%%%%%%%%%%%%%%%%%%%%%%%%%%%%%%%%%%

\clearpage
%%%%%%%%%%%%%%%%%%%%%%%%%%%%%%%%%%%%%%%%%%%%%%%%%%%%%%%%%%%%%%%%%%%

\clearpage
%\bibliography{Reference_MBT}
%merlin.mbs apsrev4-1.bst 2010-07-25 4.21a (PWD, AO, DPC) hacked
%Control: key (0)
%Control: author (0) dotless jnrlst
%Control: editor formatted (1) identically to author
%Control: production of article title (0) allowed
%Control: page (1) range
%Control: year (0) verbatim
%Control: production of eprint (0) enabled
%